\documentclass[%
 reprint,
 amsmath,amssymb,
 aps,
pra,
]{revtex4-1}

\usepackage{graphicx}
\usepackage{dcolumn}
\usepackage{bm}
\usepackage{color}
\usepackage{blindtext}
\DeclareMathOperator{\sech}{sech}
\usepackage{physics}
\usepackage{enumerate}
\newcommand{\dash}{-}%
\newcommand{\Lagr}{\mathcal{L}}

\begin{document}

\preprint{APS/123-QED}

\title{Stability of matter-wave solitons in a density-dependent gauge theory}


\author{R. J. Dingwall, and P. \"{O}hberg}
\affiliation{\\ SUPA, Institute of Photonics and Quantum Sciences,
Heriot-Watt University, Edinburgh, EH14 4AS, United Kingdom}





\date{\today}

\begin{abstract}
We consider the linear stability of chiral matter-wave solitons described by a density-dependent gauge theory. By studying the associated Bogoliubov-de Gennes equations both numerically and analytically, we find that the stability problem effectively reduces to that of the standard Gross-Pitaevskii equation, proving that the solitons are stable to linear perturbations. In addition, we formulate the stability problem in the framework of the Vakhitov-Kolokolov criterion and provide supplementary numerical simulations which illustrate the absence of instabilities when the soliton is initially perturbed.
\end{abstract}

\pacs{Valid PACS appear here}
\maketitle


\section{Introduction}
The stability of solitons in the presence of perturbations represents one of the fundamental problems in the study of solitary waves. Only solitons which are generally robust to perturbations are suitable for study in a physical setting, and by extension, implementing into potential applications in both science and industry \cite{haus_1996,ablowitz_2000}. Over the years, this topic has been studied extensively for various nonlinear models, with analysis generally falling into two frameworks: the study of small (linear) perturbations of the soliton envelope via a linear stability analysis \cite{zakharov_2012,fogel_1977,book_yang,book_spat}, or the study of additional perturbative terms in the model through a perturbation theory or variational analysis \cite{kaup_1976,bondeson_1979,kivshar_1989,elgin_1993,anderson_1983a,merterns_2010}. The motivation here is twofold: to establish the intrinsic stability of the soliton in a given model, but also to consider the effect of physically relevant perturbations which may influence or potentially damage the soliton. The latter point is of particular importance, as realistic systems are generally described by non-integrable models, in which solitons can potentially be unstable.

In the setting of nonlinear optics, described by the generalized nonlinear Schr\"{o}dinger equation, understanding the stability of solitons in the presence of perturbations has been a fundamental area of research in the design of soliton-based optical communications \cite{hasegawa_1973,book_agrawal}. Most notable is the Vakhitov-Kolokolov criterion \cite{vk_1973,weinstein_1986}, which connects the linear stability of bright solitons to two key properties: the number of negative eigenvalues in the spectral problem, and the behaviour of the power integral with respect to the propagation constant. In addition to this pioneering work, several studies have addressed the addition of perturbative terms in the model, such as but not limited to, the excitation of internal modes in non-Kerr media \cite{kivshar_1998,pelinovsky_1998,pelinovsky_1996}, the effects of third-order dispersion and self-steepening \cite{anderson_1983b,elgin_1993}, and more recently for $\mathcal{P}\mathcal{T}$-symmetric potentials \cite{PT_1,PT_2,PT_3,PT_4}, which describe media with complex refractive indices. 

In ultracold atomic gases, the linear stability framework is more commonly referred to as the Bogoliubov-de Gennes equations and plays a historic role in understanding the superfluid properties of the gas \cite{bogo_1947,book_PandS,dalfovo_1999}. Following the first experimental realisation of these ultracold gases, considerable work was centered around studying the response of the trapped condensate to small perturbations \cite{stringari_1996,edwards_1996,ruprecht_1996,jin_1996,ohberg_1997}, highlighting the collective nature of the low-lying excitations. The second generation of studies considered the case of both dark and bright solitons \cite{muryshev_1999,becker_2008,frantzeskakis_2010,perez_1998,khaykovich_2002,strecker_2002}, which in turn has lead to a number of key works, such as the interactions between trapped bright solitons \cite{nguyen_2014,martin_2007}, generation of soliton trains \cite{strecker_2002,carr_2004,nguyen_2017}, the reflection properties of bright solitons \cite{sinha_2006,lee_2006}, and understanding the snaking instability of dark solitons \cite{brand_2002,ma_2010}.

As the design of these ultracold systems become more involved, it is important to understand how the dynamics and stability of the condensate is modified, as retaining the coherent property of the gas is of vital importance for the purposes of interferometry and future atomtronic based technologies \cite{helm_2015,mcdonald_2017,seaman_2007}. In particular, the ability to simulate artificial gauge potentials in these systems \cite{ohberg_2011,goldman_2014} can lead to the emergence of vortices \cite{lin_2009} and spin-orbit coupling \cite{lin_2011}, offering new means to control and manipulate the gas. Recently, proposals have appeared which center around the engineering of density-dependent gauge potentials which feature a back-action between the matter-field and the gauge potential \cite{edmonds_2013,greschner_2014}. The condensate dynamics in these models can be greatly modified, including collective modes which violate Kohn's theorem \cite{edmonds_2015,zheng_2015,chen_2012,zhang_2012}, unconventional vortex dynamics \cite{butera_2016a,butera_2016b}, and the emergence of chiral solitons which feature non-integrable collision dynamics \cite{dingwall_2018}. Very recently, experiments have appeared in which a dynamical gauge theory was realised in trapped ion systems \cite{martinez_2016} and a density-dependent synthetic gauge field in a Bose-Einstein condensate loaded into a two-dimensional lattice \cite{clark_2018}.

In this paper, we study the linear stability of one-dimensional chiral matter-wave solitons. We begin, by briefly reviewing the physical model, in which a density-dependent gauge potential is optically engineered in a ultracold atomic gas. Then, in Section ~\ref{section_bdg}, we derive
the Bogoliubov-de Gennes equations and study the stability spectrum both numerically and analytically in Sec.~\ref{section_num} and Sec.~\ref{section_ana} respectively. Finally, in Sec.~\ref{section_VK}, we reformulate the stability problem using the Vakhitov-Kolokolov criterion, before concluding in Sec.~\ref{section_conc}.

\section{The model}
We consider the model studied in Ref.~\cite{edmonds_2013}, in which a harmonically-trapped two-level Bose-Einstein condensate is optically dressed by an external laser field. The Hamiltonian describing the system can be written as
\begin{equation}\label{Hammy}
\hat{H}=\bigg(\frac{\hat{\mathbf{p}}^{2}}{2m}+\frac{1}{2}m\omega _{\perp
}^{2}\mathbf{r}_{\mathrm{\perp }}^{2}\bigg)\otimes \mathbb{I}+\hat{H}_{
\mathrm{int}},
\end{equation} 
with the interaction matrix 
\begin{equation}
\hat{H}_{\mathrm{int}}=
 \begin{pmatrix}
  g_{11}\vert\Psi_1\vert^2+g_{12}\vert\Psi_2\vert^2 & e^{-i\phi_\ell}\frac{\hbar\Omega}{2}  \\
  e^{i\phi_\ell}\frac{\hbar\Omega}{2} & g_{12}\vert\Psi_1\vert^2+g_{22}\vert\Psi_2\vert^2
 \end{pmatrix}.
\end{equation}
The light-matter interactions are parametrised by the Rabi frequency $\Omega$ and phase $\phi_\ell(\mathbf{r})$ of the laser field, with the mean-field interactions controlled by the scattering parameters $g_{ii^\prime}=4\pi\hbar^2a_{ii^\prime}/m$, with $a_{ii^\prime}$ corresponding to the scattering lengths for collisions between atoms in state $i$ and $i^\prime$. The harmonic potential appearing in Eq.~\eqref{Hammy} is chosen to be tightly confined in the radial plane $\mathbf{r}_{\mathrm{\perp}}\left(y,z\right)$, but free along the axial $x$-axis such that the condensate dynamics is effectively one-dimensional.

By treating the mean-field interactions as a small perturbation to the laser coupling $\hbar\Omega\gg g_{ii^\prime}\vert\Psi_i\vert^2$, it can be shown in the dressed state picture \cite{edmonds_2013,ohberg_2011}, that the effective detuning induced by the interacting gas can give rise to a density-dependent gauge potential
\begin{equation}\label{vec_pot}
\begin{split}
\mathbf{A}_{\pm }&=\mathbf{A}^{(0)}\pm\mathbf{a}_{1}\vert\Psi _{\pm }(%
\mathbf{r})\vert^{2}\\
&=-\frac{\hbar }{2}\nabla \phi _{l}(\mathbf{r})\pm\frac{\nabla \phi _{l}(\mathbf{r})(g_{11}-g_{22})}{8\Omega}\vert\Psi _{\pm }(%
\mathbf{r})\vert^{2},
\end{split}
\end{equation}
where $\mathbf{A}^{(0)}$ is the single-particle vector potential and $\mathbf{a}_{1}$ controls the strength of the density-dependent gauge potential with $\pm$ indices labelling the dressed states. In the following we choose without loss of generality, one of the dressed states and drop the $\pm$ indices. By reducing to the one-dimensional picture, the dynamics is described by the Gross-Pitaevskii equation
\begin{equation}\label{EOM_normal}
i\hbar\frac{\partial\Psi}{\partial t}=\bigg[\frac{1}{2m}\left(\hat{p}-a_1\vert\Psi\vert^2\right)+a_1j(x)+g_{\mathrm{1D}}\vert\Psi\vert^2\bigg]\Psi,
\end{equation}
which contains a probability current of the form
\begin{equation}\label{current_normal}
j(x)=\frac{1}{2m}\bigg[\Psi\left(\hat{p}+a_1\vert\Psi\vert^2\right)\Psi^*-\Psi^*\left(\hat{p}-a_1\vert\Psi^*\vert^2\right)\Psi\bigg],
\end{equation}
in addition to the standard cubic nonlinearity. The strengths of the scattering parameters are given by $a_1=k_l\left(g_{11}-g_{22}\right)/\left(16\pi l^2_\perp\Omega\right)$ and $g_{\mathrm{1D}}=\left(g_{11}+g_{22}+2g_{12}\right)/\left(8\pi l^2_\perp\right)$, which are scaled by the harmonic length $l_\perp=\sqrt{\hbar/m\omega_\perp}$ and laser phase $\phi_l=k_lx$.  The interacting gauge theory described by Eq.~\eqref{EOM_normal} represents a novel nonlinear model, in which the condensate dynamics is influenced by a back-action between the matter-field and the gauge potential.

Rather than working with Eq.~\eqref{EOM_normal} directly, we will instead consider the system
\begin{equation}\label{EOM}
i\hbar\frac{\partial\psi}{\partial t}=\bigg[-\frac{\hbar^2}{2m}\partial^2_x-2a_1j^\prime(x)+g_{\mathrm{1D}}\vert\psi\vert^2\bigg]\psi,
\end{equation}
with
\begin{equation}
j^\prime(x)=\frac{\hbar}{2mi}\left(\psi^*\partial_x\psi-\psi\partial_x\psi^*\right),
\end{equation}
which is arrived at by using the nonlinear transformation
\begin{equation}\label{nonlinear_trans}
\Psi\left(x,t\right)=\psi\left(x,t\right)\mathrm{exp}\left(\frac{ia_1}{\hbar}\int^x_{\infty}dx^\prime\;\vert\psi\left(x^\prime,t\right)\vert^2\right).
\end{equation}
In the literature, Eq.~\eqref{EOM} is often referred to as a \textit{`chiral nonlinear Schr\"{o}dinger equation'}, which was originally studied in the context of one-dimensional anyons \cite{aglietti_1996}. Compared to the standard Gross-Pitaevskii equation, this model is generally \textit{non-integrable} \cite{chen_1979,nishino_1998}, and posses chiral soliton solutions which arise due to the breakdown of Galilean invariance \cite{jackiw_1997,kumar_1998}. As solitons in non-integrable models can be unstable to perturbations \cite{book_spat,book_yang}, we are naturally concerned with the stability of the soliton solutions in our model.
\subsection{Conservation laws}
The principle conservation laws underlying the chiral model are given by the integral expressions \cite{jackiw_1997}
\begin{equation}\label{norm_integral}
N=\int^{\infty}_{-\infty}dx\;\vert\psi\vert^2,
\end{equation}
\begin{equation}\label{mom_integral}
P=-i\hbar\int^{\infty}_{-\infty}dx\;\psi^*\partial_x\psi+a_1\int^{\infty}_{-\infty}dx\;\vert\psi\vert^4,
\end{equation}
and
\begin{equation} \label{ener_integral}
E=\int^{\infty}_{-\infty}dx\;\left(\frac{\hbar^2}{2m}\vert\partial_x\psi\vert^2+\frac{g_\mathrm{1D}}{2}\vert\psi\vert^4\right),
\end{equation}
which quantify the number of atoms (power integral), momentum, and energy of the condensate respectively. An important dynamical feature of the model is highlighted by Eq.~\eqref{mom_integral}, which shows that the gauge field contributes to the momentum of the condensate in a similar manner for a particle travelling in an electromagnetic field. Note, that the Hamiltonian density defined by the integrand of Eq.~\eqref{ener_integral} excludes the probability current, but does correctly reduce to Eq.~\eqref{EOM} provided Hamilton's equations are also transformed via Eq.~\eqref{nonlinear_trans}.
\subsection{Chiral solitons}
In order to derive and subsequently study the stability of the soliton solutions of Eq.~\eqref{EOM}, it will prove advantageous to work in the moving frame of the condensate as opposed to the stationary frame. To this end, we introduce the Gailiean transformation
\begin{equation}\label{trans_Galil}
\psi(x,t)=\Phi(x^\prime,t^\prime)e^{i(mvx^{\prime}+mv^{2}t^{\prime
}/2)/\hbar },
\end{equation}
where the stationary coordinates $(x,t)$ and moving coordinates $(x^\prime,t^\prime)$ are related by the translations, $x^\prime\rightarrow x-vt$ and $t^\prime\rightarrow t$, with frame velocity $v$. The dynamics of the condensate in the moving frame is then described by the equation of motion
\begin{equation}\label{EOM_mf}
i\hbar\frac{\partial\Phi}{\partial t^\prime}=\bigg[-\frac{\hbar^2}{2m}\partial^2_{x^\prime}-2a_1j^\prime(x^\prime)+\left(g_{\mathrm{1D}}-2a_1v\right)\vert\Phi\vert^2\bigg]\Phi.
\end{equation}
The introduction of the renormalised scattering parameter $\tilde{g}_{\mathrm{1D}}=g_{\mathrm{1D}}-2a_1v$, highlights that Eq.~\eqref{EOM} is not Galilean invariant, with the strength of the mean-field interactions dependent on both the magnitude and direction that the condensate is moving. For the remainder of this paper, we will explicitly drop the prime notation in the coordinates for brevity and work exclusively in the moving frame unless otherwise stated.

The bright soliton solutions of Eq.~\eqref{EOM_mf} then admit the standard form $\Phi\left(x,t\right)=\varphi_\mathrm{S}(x)e^{-i\mu t/\hbar}$, with envelope
\begin{equation}\label{brightsoliton}
\varphi_{\mathrm{S}}\left(x\right)={\frac{1}{\sqrt{2b}}}\sech\left(x
/b\right),
\end{equation}
width $b=-2\hbar^2/m\tilde{g}_{\mathrm{1D}}$, and chemical potential (propagation constant) $\mu=-m\tilde{g}^2_{\mathrm{1D}}/8\hbar^2$. In this example, each of these quantities is constrained by normalising $\varphi_\mathrm{S}$ to unity, provided $\tilde{g}_\mathrm{1D}<0$. Due to the breakdown of Galilean invariance, both the width and chemical potential of the soliton will depend on the direction of motion. The soliton solution described by Eq.~\eqref{brightsoliton} is therefore chiral, such that under appropriate conditions the soliton can either be stable or unstable in a given direction \cite{edmonds_2013}. In recent work, we have shown how the non-integrability of the model can lead to interesting collision dynamics between pairs of chiral solitons, featuring inelastic trajectories and population transfer, in addition to soliton fission and the formation of two-bounce resonance states \cite{dingwall_2018}.

\section{Bogoliubov-de Gennes equations}\label{section_bdg}
To study the linear stability of the chiral soliton, we proceed in the standard way by introducing the condensate wave function \citep{book_PandS,pelinovsky_1998}
\begin{equation}\label{excitation}
\Phi\left(x,t\right)=\left(\varphi_{\mathrm{S}}+\left(u+v\right)e^{-i\omega t}-\left(u-v\right)^*e^{i\omega t}\right)e^{-i\mu t/\hbar},
\end{equation}
in which the soliton envelope $\varphi_\mathrm{S}$ is perturbed by small-amplitude excitations, $u\left(x\right)$ and $v\left(x\right)$, with frequency $\omega$. Substituting Eq.~\eqref{excitation} into Eq.~\eqref{EOM_mf} and linearising to first order in $u\left(x\right)$ and $v\left(x\right)$, leads to the zeroth-order equation
\begin{equation}\label{zeroth}
\mathcal{L}_1\varphi_{\mathrm{S}}=0,
\end{equation}
and the pair of Bogoliubov-de Gennes equations \cite{edmonds_2015}
\begin{equation}\label{bogoeqn_1}
 \hat{L}
 \begin{pmatrix}
 u \\
 v
 \end{pmatrix}
 =\begin{pmatrix}
  2\mathcal{J}_0 & \Lagr_3\\
  \Lagr_1& 0
 \end{pmatrix}\begin{pmatrix}
 u \\
 v
 \end{pmatrix}
 =\hbar\omega\begin{pmatrix}
 u \\
 v
 \end{pmatrix},
\end{equation}
which feature the standard linear operators
\begin{equation}\label{operator_1}
\mathcal{L}_\kappa=-\frac{\hbar^2}{2m}\partial^2_x-\mu+\kappa\tilde{g}_{\mathrm{1D}}\vert\varphi_{\mathrm{S}}\vert^2,
\end{equation}
for $\kappa=\{1,3\}$, in addition to the current operator
\begin{equation}\label{operator_2}
\mathcal{J}_0=\frac{ia_1\hbar}{m}\left[\vert\varphi_{\mathrm{S}}\vert^2\partial_x-\frac{1}{2}\partial_x\vert\varphi_{\mathrm{S}}\vert^2\right].
\end{equation} 
Together, Eqs.~\eqref{zeroth} and \eqref{bogoeqn_1} describe the perturbation dynamics of the soliton, with the stability properties determined by the nature of the eigenvalues, or `stability spectrum', of the Bogoliubov-de Gennes equations. Note, that the linearised operator $\hat{L}$ is not self-adjoint (see Eq.~\eqref{bogoeqn_1_adj}), even in the standard case with $a_1=0$.

A key feature of these equations is highlighted by the property that in the moving frame, the current operator does not explicitly couple to the envelope of the soliton, but does couple to the excitations around it. In turn, this leads to the zeroth-order equation for the stationary soliton being described by the \textit{integrable} Gross-Pitaevskii equation, despite Eq.~\eqref{EOM} being generally \textit{non-integrable}. Therefore, it is expected that the spectrum of excitations around the chiral soliton will be similar to that of integrable models, with only their form modified slightly due to the coupling of the current operator, as per Eq.~\eqref{operator_2}. This proposition will be a key underlying point in the analysis to follow.

We can also conclude several additional properties of the excitations by studying the matrix
\begin{equation}\label{bogoeqn_2}
\hat{L}= 
 \begin{pmatrix}
  \left(ia_1\hbar/mb\right)\sech^2\left(x/b\right)\partial_x-iW & \Lagr_0+3V\\
  \Lagr_0+V & 0 
 \end{pmatrix},
\end{equation}
which is obtained by substituting the soliton solution into Eq.~\eqref{bogoeqn_1}.
Appearing in Eq.~\eqref{bogoeqn_2} are two potential functions
\begin{equation}
V(x)=\frac{\tilde{g}}{2b}\sech^2(x/b),
\end{equation}
and
\begin{equation}
iW(x)=i\frac{\tilde{g}_\mathrm{1D}}{2b}\frac{a_1}{\hbar}\tanh(x/b)\sech^2(x/b),
\end{equation}
which are a standard reflectionless potential \cite{lekner_2007,var_1} and a gain-loss distribution for the excitations respectively \cite{PT_1,PT_2,PT_3,PT_4}. Together, they form a modified (hyperbolic) Scarf-II potential with the following properties \cite{PT_1}:
\begin{enumerate}[(i)]
\item Bounded with $V(x)\leq 0$, $\vert W(x)\vert\leq \tilde{g}_\mathrm{1D}a_1/\left(3\sqrt{3}b\hbar\right)$.
\item Convergent with $x\rightarrow\infty$, $V(x)$ and $W(x)\rightarrow 0$.
\item Not self-adjoint $\left(iW(x)\right)^\dag\neq iW(x)$ (see Eq.~\eqref{operator_adjoint}).
\item $\mathcal{P}\mathcal{T}$-symmetric, $\hat{\mathcal{P}}\hat{\mathcal{T}}\left(V(x)+iW(x)\right)=V(x)+iW(x)$.
\end{enumerate}
These properties highlight that the soliton acts as a complex effective potential for the excitations, with bound states and scattering states supported for the attractive potential $V(x)$, and the gains and losses of the excitations balanced by the symmetry of the imaginary potential $\int^\infty_{-\infty}dx\;W(x)=0$. As this potential is not self-adjoint, the stability spectrum for the excitations could potentially contain complex eigenvalues, in addition to a $\mathcal{P}\mathcal{T}$-symmetry breaking point featuring exceptional points \cite{heiss_2012}.

\section{Numerical Results}\label{section_num}
\begin{figure}[t]
\includegraphics[width=8.8cm]{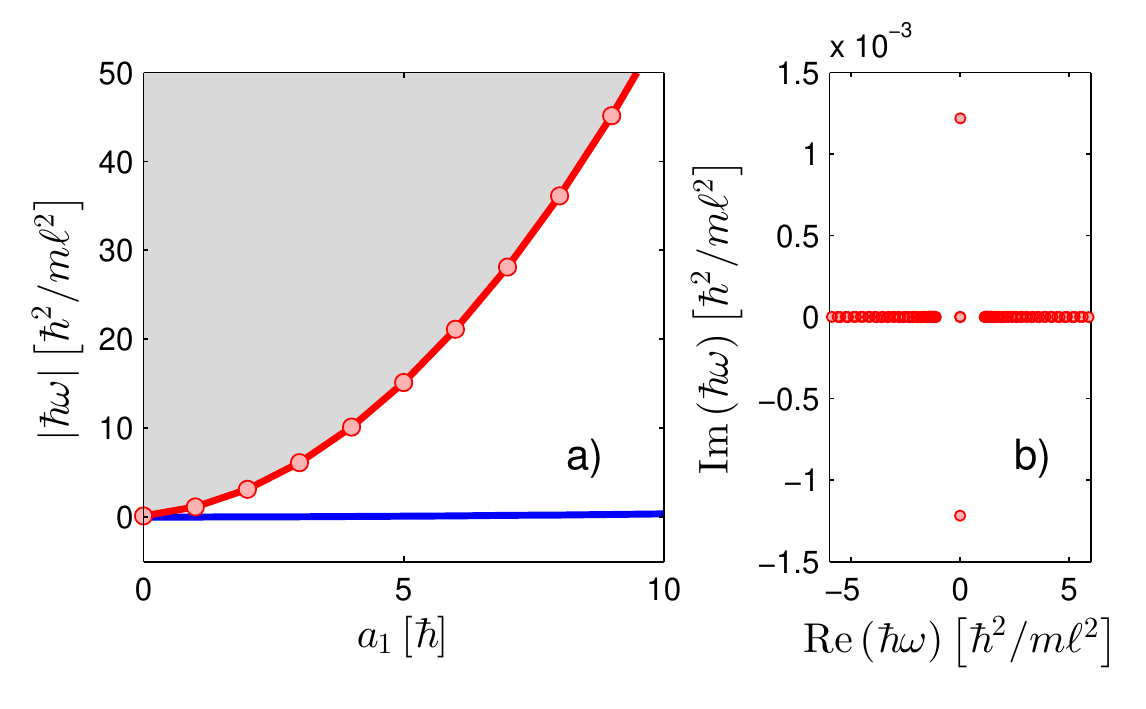}
\caption{(colour online). (a) Numerically obtained eigenvalue spectrum of the Bogoliubov-de Gennes equations with discrete states (blue) and continuous states (grey). The band edge of the continuous spectrum is highlighted in red, for both numerical (solid) and analytical (dots) results.  The soliton parameters are taken as $g_{\mathrm{1D}}m\ell /\hbar ^{2}=-1$, and $vm\ell /\hbar =1$. (b) Subset of (a) taken at $a_{1}/\hbar =1$.\\~\\}
\label{image_bogo_eigs}

\includegraphics[width=8.8cm]{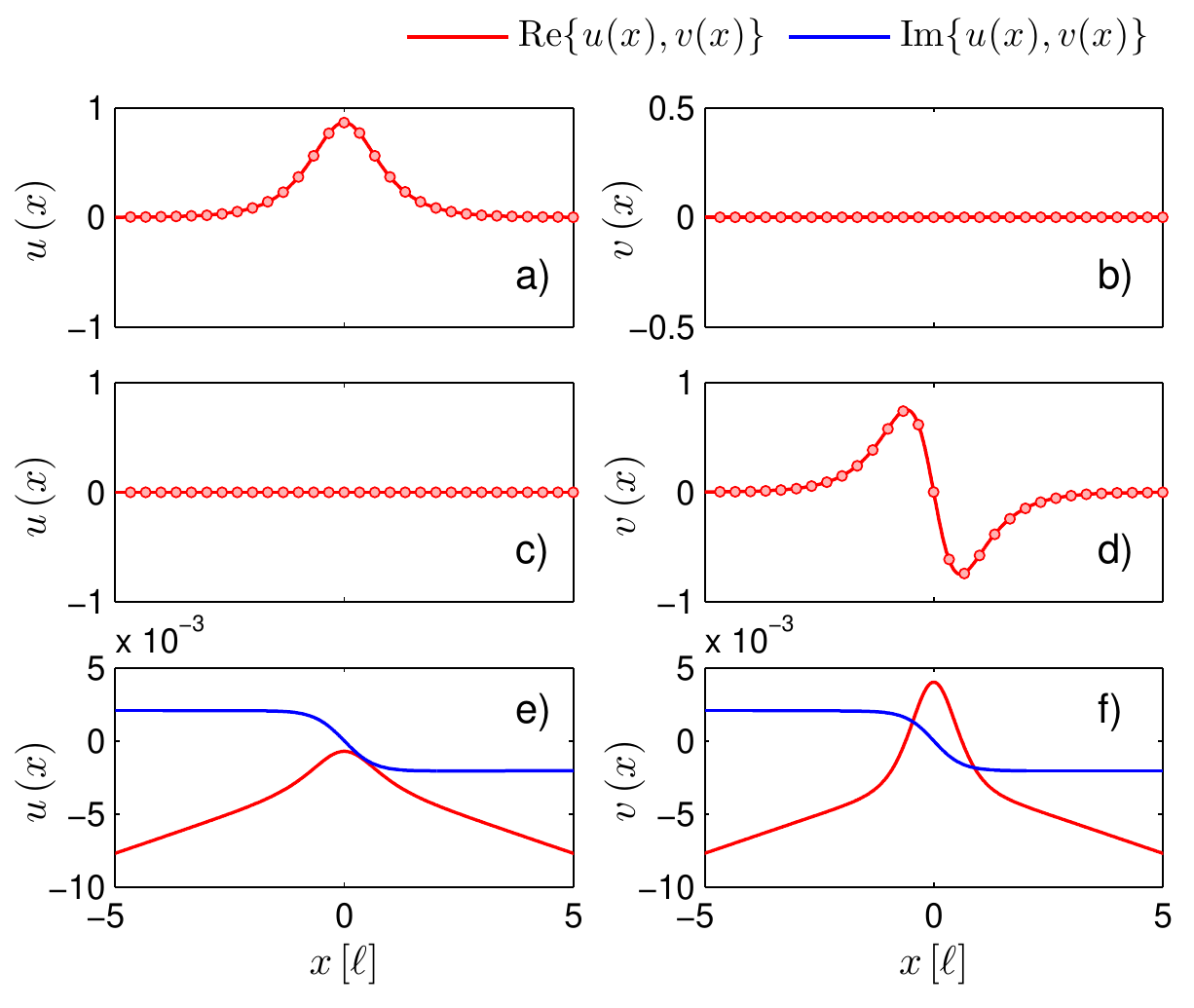}
\caption{(colour online). Bogoliubov-de Gennes eigenvectors $\left(u\enspace v\right)^\mathrm{T}$ for $g_{\mathrm{1D}}m\ell /\hbar ^{2}=-1$, $vm\ell /\hbar =1$, and $a_{1}/\hbar =1$. Pictured are the degenerate bound states, (a-b) and (c-d), corresponding to the discrete spectrum and the first continuous state (e-f). All eigenvectors are scaled to units of $\ell^{-1/2}$ for both numerical (solid-line) and analytical (dots) results. \\}
\label{image_bogo_vecs}
\end{figure}

With the Bogoliubov de-Gennes equations derived and their properties reviewed, we can now proceed in solving for the stability spectrum of the soliton. To achieve this, we first consider a numerical solution, in which Eq.~\eqref{bogoeqn_1} is discretised with periodic boundary conditions and subsequently solved using a sparse eigenvalue solver. The resulting eigenvalues and eigenvectors are shown in Figure~\ref{image_bogo_eigs} and Fig.~\ref{image_bogo_vecs} respectively.

As expected, we find that the eigenspectrum retains several characteristic features commonly encountered in integrable models \cite{book_yang}. The eigenvalues consist of the union of two sets: a continuous spectrum with two symmetric branches each gapped from the origin by $\vert\hbar\omega\vert=\vert\mu\vert$, and a discrete spectrum with one pair of eigenvalues located at $\hbar\omega=0$ and another pair displaced from the origin by a small imaginary component. At first glance, this pair of imaginary eigenvalues would indicate an instability mode where the soliton state can collapse. However, this component arises instead due to the discreteness of the numerical analysis and subsequently vanishes in the continuum limit (see Appendix~\ref{appen_eigs}). The eigenvalues of Eq.~\eqref{bogoeqn_1} are therefore entirely real with a four-fold degenerate eigenvalue at $\hbar\omega=0$, in an identical manner as for the Gross-Pitaevskii equation. In reference to the $\mathcal{P}\mathcal{T}$-symmetry breaking proposed earlier, we find that the eigenvalues are consistently real in the parameter space $\tilde{g}_\mathrm{1D}<0$, indicating the absence of a $\mathcal{P}\mathcal{T}$-symmetry breaking point for the regime in which the soliton solution is valid. Therefore we may conclude that in \textit{free space}, the chiral soliton is stable to linear perturbations. 

The eigenvectors of Eq.~\eqref{bogoeqn_1} are also consistent with that of integrable models, with the discrete spectrum corresponding to localised real-valued solutions in the vicinity of the soliton (Fig.~\ref{image_bogo_vecs}.(a-d)), while the continuous states are complex-valued and generally oscillatory at $x=\pm\infty$ (Fig.~\ref{image_bogo_vecs}.(e-f)). In fact, the discrete states pictured are exactly the same as for the Bogoliubov-de Gennes equations associated with the Gross-Pitaevskii equation, with the first and second states taking the form of the envelope of the soliton and its derivative respectively. This similarity, or rather, the invariance of the form of the discrete states in the presence of the current operator, will become clear from the analytical results in the next section.

In obtaining these numerical results, the proposition that the stability spectrum of the chiral soliton is similar to that of integrable models is validated. As such, the interacting gauge theory represents a non-integrable model in which the soliton solutions are stable to linear perturbations. The absence of instability modes in the model will be revisited in the numerical simulations presented in Sec.~\ref{section_numerics}.

\section{Analytical Results}\label{section_ana}

For several integrable models, the spectrum of excitations around the soliton solution can be derived analytically \cite{fogel_1977,sachs_1983,kaup_1990}. This is generally achieved either through a connection to the squared eigenfunctions of the associated eigenvalue problem \cite{yang_2000}, or in some cases, by direct methods. As our model is generally non-integrable, it is unclear whether the former method would be applicable. However, using the numerical results obtained previously as a basis, we can at the very least obtain expressions for the discrete spectrum using traditional methods.
\subsection{Discrete Spectrum - Bound States}
As was demonstrated previously, the discrete spectrum of the excitations correspond to a set of bound states which are a zero-eigenvalue solution of Eq.~\eqref{bogoeqn_1}. For this reason, the Bogoliubov-de Gennes equations can be written as
\begin{equation}\label{f_eqn}
\mathcal{L}_1u=0,
\end{equation}
and
\begin{equation}\label{g_eqn}
\mathcal{L}_3v=-2\mathcal{J}_0u.
\end{equation}
The above set of differential equations can be solved analytically using either a hypergeometric series approach \cite{book_flugge,book_LL,book_DJ}, or by operator methods \cite{book_super} (see Appendix~\ref{appen_hyper} for details on the former). Then, by denoting $\zeta_n=\begin{pmatrix}u&v\end{pmatrix}^\mathrm{T}$ for $n\in\mathbb{Z}^+$, we find the pair of zero-eigenvalue solutions
\begin{equation}
 \zeta_1= 
 \begin{pmatrix}
c_1\sech\chi \\
 0
 \end{pmatrix},
\end{equation}
and
\begin{equation}
 \zeta_2= 
 \begin{pmatrix}
0 \\
 c_4\sech\chi\tanh\chi
 \end{pmatrix},
\end{equation}
where $\chi=x/b$, and $c_1$ and $c_4$ are arbitrary constants to be determined.
Due to the four-fold degeneracy of the zero-eigenvalue, two linearly-independent solutions, $\zeta_3=2\left(\zeta_{3,-}+\zeta_{3,+}\right)$ and $\zeta_4=2\left(\zeta_{4,-}+\zeta_{4,+}\right)$, are also required for completeness, where
\begin{equation}
 \zeta_{3,-}=
 c_2\sech\chi
 \begin{pmatrix}
 \chi \\
 ia_1/\left(2\hbar\right)
 \end{pmatrix},
\end{equation}
\begin{equation}
 \zeta_{3,+}=
 c_2
 \begin{pmatrix}
 \sinh\chi \\
-\left(2ia_1/\hbar\right)\sech\chi\{\chi\tanh\chi-3/4\}
 \end{pmatrix},
\end{equation}
\begin{equation}
 \zeta_{4,-}=
 \begin{pmatrix}
0 \\
 c_3\sech\chi\{\chi\tanh\chi-1\}
 \end{pmatrix},
\end{equation}
and
\begin{equation}
 \zeta_{4,+}=
 \begin{pmatrix}
 0\\
 \left(c_3/3\right)\cosh\chi
 \end{pmatrix},
\end{equation}
are a set of generalized eigenvectors which satisfy the eigenrelations, $\hat{L}\zeta_{3,\pm}=\pm 2\zeta_2$ and $\hat{L}\zeta_{4,\pm}=\pm2\zeta_1$, up to a constant \cite{kaup_1976,kaup_1990}. It is straightforward to verify that the solutions form a linearly independent set by calculating the Wronskian
\begin{equation}
W\left(\zeta_1,\zeta_2\right)=\mathrm{det}\begin{vmatrix}
\sech\chi & 0\\
0 & \tanh\chi\sech\chi
\end{vmatrix}\neq 0,
\end{equation}
and likewise for the remaining solutions.

Together, these states compose the discrete spectrum of Eq.~\eqref{bogoeqn_1}, corresponding each to small variations of the soliton with respect to its four free parameters: phase, position, velocity, and chemical potential. As an example, by expanding the soliton solution around a small variation of the velocity $\delta v$, such that $b\rightarrow b+\delta b$, one finds to first-order that
\begin{equation}
\begin{split}
\sech\chi\;&e^{i\delta vx}=\sech\chi\\
&+i\delta vx\sech\chi\left(1-2ia_1\tanh\chi\right)+\mathcal{O}(x)^2,
\end{split}
\end{equation}
which is captured by $\zeta_3$ with reference to Eq.~\eqref{excitation}. Surprisingly, this set of discrete states is identical to that of the Bogoliubov-de Gennes equations for the Gross-Pitaevskii equation, except for $\zeta_3$ which features current-dependent terms due to the width of the soliton being defined in part by its velocity. This similarity becomes clear when obtaining the solutions, as one finds that the soliton envelope is a zero eigenvalue solution of the operator $\Lagr_1\varphi_\mathrm{S}=0$ and the linearised current operator $\mathcal{J}_0\varphi_\mathrm{S}=0$. Therefore, the eigenvalue problem for both $\zeta_1$ and $\zeta_2$ effectively reduces to the standard case, which naturally leads to the form of the solutions pictured in Fig.~\ref{image_bogo_vecs}.

Defining the inner product between any pair of solutions as
\begin{equation}\label{inner_product}
\langle f_n,f_{n^\prime}\rangle\equiv\int^\infty_{-\infty}dx\;f_n^\dag f_{n^\prime},
\end{equation}
together with the normalisation constraint defined by Eq.~\eqref{norm_integral}, leads to the values of $\vert c_1\vert^2=1/\left(2b\right)$ and $\vert c_4\vert^2=3/\left(2b\right)$ for the arbitrary constants. Note, that since $\zeta_3$ and $\zeta_4$ are not square-integrable solutions, we are not required to compute inner-products for the remaining constants as these states are not physical.

With the form of the solutions now reduced, we are now in a position to compare our analytical results to the numerical ones obtained earlier. In Fig.~\ref{image_bogo_vecs}, we find exact agreement between the analytical (dots) and numerical (solid-line) results for the degenerate bound states $\zeta_1$ and $\zeta_2$. This reinforces the statement that the discrete eigenvalues in the chiral model are four-fold degenerate at $\hbar\omega=0$, with the imaginary component in the numerical analysis attributed to numerical artifacts. Therefore we may conclude, that due to the consistent results obtained from both methods, that the chiral soliton is stable to linear perturbations.
\subsection{Discrete Spectrum - Adjoint Bound States}
In addition to the standard eigenvalue problem, we can also solve the corresponding adjoint problem using the same analytical techniques. Although this system does not have any physical relevance, the equivalence of both eigenvalue problems will be a key property which we will exploit when deriving the Vakihitov-Kolokolov criterion in the next section.

For the sub-space of square-integrable solutions of $\hat{L}$ with the inner-product defined by Eq.~\eqref{inner_product}, we write the adjoint operator
\begin{equation}\label{bogoeqn_1_adj}
 \hat{L}^\dag
 =\begin{pmatrix}
  2\mathcal{J}_0^\dag & \Lagr_1 \\
  \Lagr_3 & 0
 \end{pmatrix},
\end{equation}
whose right-eigenvectors are the adjoint of the left-eigenvectors of $\hat{L}$. The adjoint of the linearised current-operator appearing in Eq.~\eqref{bogoeqn_1_adj} is given by
\begin{equation}\label{operator_adjoint}
\mathcal{J}_0^\dag=\frac{ia_1\hbar}{m}\left[\vert\varphi_{\mathrm{S}}\vert^2\partial_x+\frac{3}{2}\partial_x\vert\varphi_{\mathrm{S}}\vert^2\right].
\end{equation} 
Together, Eqs.~\eqref{bogoeqn_1_adj} and \eqref{operator_adjoint} explicitly highlight that $\hat{L}$, as mentioned previously, is not self-adjoint.

Using the same methodology as before, we denote the left-eigenvectors of $\hat{L}$ as $\vartheta_n=\begin{pmatrix}u &v\end{pmatrix}$ and find the adjoint solutions
\begin{equation}
\vartheta_1=
 \begin{pmatrix}
 0 \\
 c_5\sech\chi
 \end{pmatrix}^\dag,
\end{equation}
and
\begin{equation}
\vartheta_2=
 c_8\sech\chi
 \begin{pmatrix}
 \tanh\chi \\
 -\left(ia_1/\hbar\right)\sech^2\chi
 \end{pmatrix}^\dag.
\end{equation}
Again, due to the degeneracy of the zero eigenvalue, two additional solutions,
$\vartheta_3=2(\vartheta_{3,-}+\vartheta_{3,+})$ and $\vartheta_4=2(\vartheta_{4,-}+\vartheta_{4,+})$, are also required for completeness, where
\begin{equation}
\vartheta_{3,-}=
\begin{pmatrix}
 0 \\
c_6\chi\sech\chi
 \end{pmatrix}^\dag,
\end{equation}
\begin{equation}
\vartheta_{3,+}=
\begin{pmatrix}
 0 \\
c_6\sinh\chi
 \end{pmatrix}^\dag,
\end{equation}
\begin{equation}
\vartheta_{4,-}=
c_7\sech\chi
\begin{pmatrix}
 \chi\tanh\chi-1 \\
-\left(ia_1/\hbar\right)\{\chi\sech^2\chi-\tanh\chi\}
 \end{pmatrix}^\dag,
\end{equation}
and
\begin{equation}
\vartheta_{4,+}=\frac{c_7}{3}
\begin{pmatrix}
 \cosh\chi\\
\left(2ia_1/\hbar\right)\chi\sech\chi
 \end{pmatrix}^\dag,
\end{equation}
are the set of generalized eigenvectors satisfying the eigenrelations, $L^\dag\vartheta_{3,\pm}=\pm2\vartheta_2$ and $L^\dag\vartheta_{4,\pm}=\pm2\vartheta_1$, up to a constant.
\subsection{Continuous States}
For the continuous states, we are unfortunately unable to derive a closed-form expression due to the complexity of the eigenvalue problem. Although we may be able to obtain these solutions using a power series method \cite{kovrizhin_2001,yan_1996}, the complicated nature of the calculation presents little benefit for the knowledge gained, since the continuous states are irrelevant for addressing the stability of the soliton. However, we can still obtain an expression for the eigenvalues using an asymptotic approach as follows.

The key point to note, is that for large distances away from the centre of the soliton, the continuous states are generally oscillatory such that they can then be written as plane waves of the form
\begin{equation}\label{bogoeqn_3}
\lim_{x\to\pm\infty} 
 \begin{pmatrix}
 u \\
 v
 \end{pmatrix}
 \sim\pm\begin{pmatrix}
 e^{iqx} \\
 e^{iqx}
 \end{pmatrix},
\end{equation}
with wave-number $q$. Then, by substituting Eq.~\eqref{bogoeqn_3} into the asymptotically reduced form of Eq.~\eqref{bogoeqn_1}, one finds the continuous-eigenvalue expression
\begin{equation}\label{eigs_cont}
\hbar\omega_c\sim\pm\left(\frac{\hbar^2q^2}{2m}-\mu\right),
\end{equation}
which as expected, is simply a free-particle dispersion relation gapped by the chemical potential of the soliton. In Fig. \ref{image_bogo_eigs}(a), we compare both the numerical (red solid-line) and analytic (red dots) values for the continuous-state band edge $(q=0)$ which as shown, is in exact agreement.

\section{Vakhitov-Kolokolov Criterion}\label{section_VK}
As the chiral solitons present in our model are solutions to a generalized Gross-Pitaevskii equation with a real positive envelope, we can also establish their stability properties using the Vakhitov-Kolokolov criterion \cite{book_yang,vk_1973}. For the standard Gross-Pitaevskii, the stability criterion requires:
\begin{enumerate}[(i)]
\item The eigenspectrum of the operators $\Lagr_1$ and $\Lagr_3$, should contain at most only a single negative eigenvalue.
\item The slope of the power integral, $dN/d\mu\geq 0$, should be non-negative for $\mu<0$.
\end{enumerate}

The first condition is straightforward to verify by noting that both operators have a positive continuous spectrum $\hbar\omega=\left[-\mu,\infty\right)$, and a discrete spectrum defined by the eigenrelations, 
\begin{equation}
\Lagr_1\varphi_\mathrm{S}=0,
\end{equation}
and
\begin{equation}
\Lagr_3\left(\partial_x\varphi_\mathrm{S}\right)=0\quad\quad\Lagr_3\left(\varphi_\mathrm{S}\right)^2=3\mu\left(\varphi_\mathrm{S}\right)^2.
\end{equation} 
The second condition can also be proven easily by direct integration, provided the particle number (power) is correctly posed \cite{zakharov_2012}. This criterion directly follows from the definiteness of the operators in the eigenvalue problem and therefore requires us to extend the Vakhitov-Kolokolov analysis to include the linearised current operator as follows.

As the eigenvalues of $\hat{L}$ and $\hat{L}^\dagger$ are conjugate to each other, we can without loss of generality consider the spectral properties of either system. To this end, it will prove advantageous to work in the adjoint picture, as the following stability analysis is simpler while exploiting the property that the operators $\Lagr_1$ and $\Lagr_3$ are self-adjoint.

We start, by restating the adjoint eigenvalue problem
\begin{equation}\label{vak_1}
2\mathcal{J}_0^\dagger u+\mathcal{L}_1 v=\hbar\omega^*u,
\end{equation}
and
\begin{equation}\label{vak_2}
\mathcal{L}_3 u=\hbar\omega^*v.
\end{equation}
Taking the inner-product of Eq.~\eqref{vak_1} with $\varphi_{\mathrm{S}}\left(x\right)$ leads to the expression
\begin{equation}
2\langle\varphi_{\mathrm{S}},\mathcal{J}_0^\dagger u\rangle+\langle\varphi_{\mathrm{S}},\mathcal{L}_1v\rangle=\hbar\omega^*\langle\varphi_{\mathrm{S}},u\rangle.
\end{equation}
As $\mathcal{L}_1$ is self-adjoint with $\mathcal{L}_1\varphi_{\mathrm{S}}=0$, the above expression is true $\forall\omega^*$, provided the orthogonality condition 
\begin{equation}\label{ortho_cond}
\langle\varphi_{\mathrm{S}},\mathcal{J}_0^\dagger u\rangle=\langle\varphi_{\mathrm{S}},u\rangle=0,
\end{equation}
is satisfied. Therefore, for the non-zero eigenvalues in the stability analysis, we may restrict ourselves to the function space
\begin{equation}\label{space}
\mathcal{S}\equiv\{\nu\left(x\right):\langle\varphi_{\mathrm{S}},\mathcal{J}_0^\dagger u\rangle=\langle\varphi_{\mathrm{S}},u\rangle=0\},
\end{equation}
where the inverse operators $\Lagr_1^{-1}$ and $\Lagr_3^{-1}$ are definable. 

Returning to Eqs.~\eqref{vak_1} and \eqref{vak_2}, we can now proceed in constructing the stability criterion for the chiral soliton by combining both equations into the fourth-order equation
\begin{equation}\label{vak_4}
\mathcal{L}_1\mathcal{L}_3\nu=\left(\hbar\omega^*\right)^2\nu-2\hbar\omega^*\mathcal{J}_0^\dagger\nu
\end{equation}
in the function space $\mathcal{S}$. Multiplying Eq.~\eqref{vak_4} by $\mathcal{L}^{-1}_1$, taking the inner product with respect to $\nu(x)$, and completing the square leads to the expression
\begin{equation}\label{stability_cond}
\left(\hbar\omega^*\right)^2=\frac{\gamma}{\alpha}+\frac{2\beta^2}{\alpha^2}\pm\frac{\beta}{\alpha}\left(\frac{\beta^2}{\alpha^2}+\frac{\gamma}{\alpha}\right)^{1/2},
\end{equation}
with $\alpha=\langle \nu,\Lagr^{-1}_1\nu\rangle$, $\beta=\langle \nu,\Lagr^{-1}_1\mathcal{J}^\dag_0\nu\rangle$, and $\gamma=\langle \nu,\Lagr_3\nu\rangle$. The condition of stability is now set by requiring that the right-hand side of Eq.~\eqref{stability_cond} be non-negative, such that $\hbar\omega^*$ by extension is real. Otherwise for negative values, $\hbar\omega^*$ would be imaginary, thereby indicating an instability. As both $\Lagr_1$ and $\Lagr_3$ (and their inverses) are known to be positive definite in the space $\mathcal{S}$ \cite{yang_2000}, the standard term $\gamma/\alpha$ will not need to be considered in our analysis. Instead, the stability of the soliton will be resolved by studying the definiteness of $\Lagr^{-1}_1\mathcal{J}^\dag_0$.

Despite being a trivial reduction of the problem, the non-negativeness of Eq.~\eqref{stability_cond} can be guaranteed if $\mathcal{J}_0^\dagger$ is nilpotent in the domain considered, i.e. $\mathcal{J}_0^\dagger \nu=0$, such that all the eigenvalues of $\mathcal{J}_0^\dagger$ are zero. This can be proven by directly solving the eigenvalue problem
\begin{equation}\label{eigen_cond}
\mathcal{J}_0^\dagger\nu=\lambda\nu,
\end{equation}
with eigenvalue $\lambda$ and eigenfunction $\nu$. The general solution of Eq.~\eqref{eigen_cond} can be readily found,
\begin{equation}
\nu=\mathcal{C}\cosh^3\chi\enspace\mathrm{exp}\left[-i\epsilon\lambda\left(\chi+\frac{1}{2}\sinh(2\chi)\right)\right],
\end{equation}
where $\epsilon=mb^2/a_1\hbar$, and $\mathcal{C}$ is an arbitrary constant. For a ring domain of length (circumference) $L$ with periodic boundary conditions $\nu\left(-L/2\right)=\nu\left(L/2\right)$, the eigenvalues form a continuous spectrum 
\begin{equation}
\lambda=\frac{2\pi\sigma}{\epsilon\left(L/b+\sinh(L/b)\right)},
\end{equation}
with $\sigma=0,\pm 1,\pm 2,\ldots$ . Then in the combined limit where $L\rightarrow\infty$ and $\sigma\rightarrow\pm\infty$, the eigenvalues of Eq.~\eqref{eigen_cond} coalesce at $\lambda=0$, highlighting that for the ring domain considered in our numerics, the adjoint current operator $\mathcal{J}_0^\dagger$ is nilpotent. Therefore, the stability condition, Eq.~\eqref{stability_cond}, reduces to the form encountered for the standard Gross-Pitaevskii equation, from which the Vakhitov-Kolokolov criterion is known to be satisfied \cite{book_yang,vk_1973}.

\subsection{Numerics}\label{section_numerics}
\begin{figure}[t]
\includegraphics[width=9.0cm]{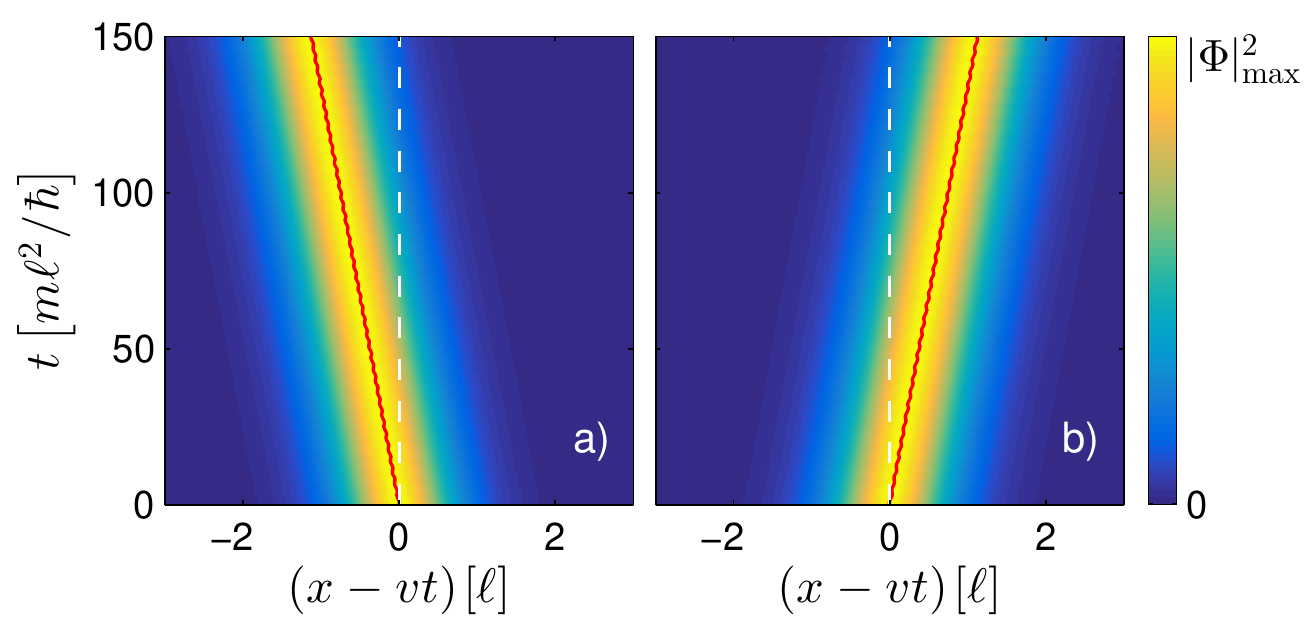}
\caption{(colour online). Propagation of a chiral soliton, in the \textit{moving frame}, whose initial envelope is perturbed due to a change in number of atoms. Shown, are the predicted trajectories from the variational equations (red-solid) in comparison to the full numerics (colour) and the unperturbed case (white-dash). The soliton parameters are $g_{\mathrm{1D}}m\ell /\hbar ^{2}=-1$, $vm\ell /\hbar =1$, and $a_{1}/\hbar =1$, with the mismatch parameters $\Delta\varphi=+0.01$ (a), and $\Delta\varphi=-0.01$ (b).}
\label{image_VK_sim}
\end{figure}

In addition to the results obtained analytically, we also consider a set of numerical simulations which illustrate the stability of the chiral soliton under the influence of a perturbation. To achieve this, we follow the standard numerical scheme in which the initial number of atoms (power) of the soliton differs from the exact solution and observe whether the soliton collapses or retains its shape \cite{book_spat,book_yang}. As such, we define the perturbed soliton state as
\begin{equation}
\varphi_\Delta=\varphi_{\mathrm{S}}\left(1+\Delta\varphi\right),
\end{equation}
and show two examples of the pertubation dynamics in Fig.~\ref{image_VK_sim}, each for a different sign of the perturbation parameter $\Delta\varphi$. 

In both cases, the soliton maintains its shape over the course of the simulation and does not collapse, disperse, or oscillate due to the excitation of an internal mode \cite{book_spat,pelinovsky_1996}. Instead, the soliton emits a small (non-visible) amount of radiation and decays to the stable low-amplitude solution, in a similar manner to solitons of the Gross-Pitaevskii equation. However, as the initial width of the soliton changes due to the perturbation, with ${g}_\mathrm{1D}$ and $a_1$ fixed, the velocity of the perturbed soliton will differ from the frame velocity set by Eq.~\eqref{trans_Galil}. This results in the soliton drifting in the moving frame, with the direction controlled by the sign of the perturbation. This effect is not captured by the stability spectrum of the Bogoliubov-de Gennes, due to it being a higher-order effect which is neglected in the linearisation. Instead, we are required to consider an alternative framework to explain the presence of the soliton drift.

\subsection{Variational equations}
To quantify the drift of the soliton due to the action of the perturbation, a variational approach can be employed to derive a set of coupled equations which effectively describe the soliton dynamics \cite{perez_1997}. Although we will not be able to solve these equations analytically, their numerical solutions will provide sufficient illustrations of the perturbation dynamics to compare with the results pictured in Fig.~\ref{image_VK_sim}.

We begin by writing the Lagrangian density \cite{kumar_1998},
\begin{equation}\label{var_Lagr}
\begin{split}
\Lagr&=\frac{i\hbar}{2}\left(\psi\partial_t\psi^*-\psi^*\partial_t\psi\right)+\frac{\hbar^2}{2m}\vert\partial_x\vert^2+\frac{g_{\mathrm{1D}}}{2}\vert\psi\vert^4\\
&+a_1\vert\psi\vert^2\frac{d}{dt}\int^x_{-\infty}dy\;\vert\psi\left(y,t\right)\vert^2,
\end{split}
\end{equation}
which is written in the \textit{stationary} frame of the soliton, with stationary coordinates $\left(x,t\right)$. To accurately describe the perturbation dynamics, we choose a general variational ansatz of the form \cite{var_1,var_2}
\begin{equation}\label{var_ansatz}
\psi\left(x,t\right)\equiv a\sech\left(\left(x-\xi\right)/b\right)e^{iS},
\end{equation}
in which the width and position of the soliton envelope can vary dynamically through the spatially-varying phase
\begin{equation}
S\left(x,t\right)\equiv k\left(x-\xi\right)+w\left(x-\xi\right)^2+\phi.
\end{equation}
Here, $a(t)$, $b(t)$, $\xi(t)$, $k(t)$, $w(t)$, and $\phi(t)$ are time-dependent variational parameters corresponding to the amplitude, width, centre-of-mass coordinate, velocity, curvature, and absolute phase of the soliton. As the form of Eq.~\eqref{var_ansatz} explicitly assumes that the shape and particle number of the soliton is conserved, the interplay of radiation will therefore be absent in the analysis. However, as we will demonstrate, provided the magnitude of the perturbation is kept small, this discrepancy will not have significant implications.

Substituting Eq.~\eqref{var_ansatz} into Eq.~\eqref{var_Lagr} and minimizing the corresponding action functional leads to the set of coupled differential equations
\begin{equation}\label{xi_eqn}
m\ddot{\xi}=\frac{Na_1\dot{b}}{3b^2},
\end{equation}
\begin{equation}\label{b_eqn}
\frac{\pi^2}{12}m\ddot{b}=\frac{\left(g-2a_1\hbar k/m\right)N}{6b^2}+\frac{\hbar^2}{3mb^3},
\end{equation}
and
\begin{equation}\label{k_eqn}
k=\frac{m\dot{\xi}}{\hbar}+\frac{Na_1}{3\hbar b},
\end{equation}
which collectively describe the motion of the soliton. The source of the drift is now clear from the coupling between Eqs.~\eqref{xi_eqn} and \eqref{b_eqn}; that a time-dependent variation of the soliton's width, induced by a perturbation, can lead to a change in the soliton's centre-of-mass proportional to the strength of the gauge field. The trajectory of the soliton is then set by Eq.~\eqref{k_eqn}, which in a consistent manner to Eq.~\eqref{mom_integral}, contains an additional contribution from the gauge field. Therefore, for either an increase or decrease in the particle number, it is expected that the soliton will drift in the moving frame.

To illustrate the above reasoning, we solve the set of differential equations numerically using a fourth-order Runge-Kutta method and plot the predicted soliton trajectories (red solid-line) in Fig.~\ref{image_VK_sim}. Both the direction and magnitude of the drift is captured correctly by the variational equations and therefore validates that the drift of the soliton arises due to how the initial state is prepared. In addition, these results show that in this weak perturbation regime, the emission of radiation from the soliton plays no significant role in the dynamics. However, its absence in the variational description does lead to inconsistencies, as demonstrated by the presence of small-amplitude oscillations in the predicted trajectories which persist indefinitely.

To conclude, although we cannot strictly say the soliton is stable due to the presence of the drift, we stress that it is a manageable feature which does not destroy or damage the envelope of the soliton. Therefore, we may view the soliton as \textit{effectively stable}, with the absence of the traditional instability mechanisms consistent with the Bogoliubov-de Gennes analysis.
\section{Conclusion}\label{section_conc}
In this paper, we have demonstrated the linear stability of chiral matter-wave solitons in an interacting gauge theory. Despite being described by a non-integrable model, we found that the stability spectrum of the soliton reduces to the standard integrable case, with entirely real eigenvalues and the absence of instability modes. This was then further understood by studying the Vakhitov-Kolokolov criterion, which highlighted that the linearised current operator was nilpotent in the numerical domain and therefore does not contribute to dynamical instabilities.

The drift of the soliton due to the presence of a perturbation represents an interesting property of the chiral model. By generalising the study to a broader class of perturbations \cite{biswas_2009}, several questions are inspired not only in regards to the stability, but also to features which could be exploited in order to control the soliton. For example, could a perturbation be designed which when applied continuously, enables the soliton to accelerate or decelerate with minimal radiation losses? These questions, together with the linear stability properties concluded in this work, offers a promising candidate for practical transport dynamics in atomtronic systems \cite{seaman_2007,amico_2017}, where retaining the coherent properties of the gas can be an important factor.
\section*{Acknowledgements}

The authors would like to thank M. J. Edmonds, J. L. Helm and B. A. Malomed for helpful discussions. R.J.D acknowledges support from EPSRC CM-CDT Grant No. EP/L015110/1, and P.\"{O} acknowledges support from EPSRC grant No. EP/M024636/1.

\appendix
\section{Convergence of Eigenvalues}\label{appen_eigs}
In Sec.~\ref{section_num}, we claim that the imaginary component of the discrete eigenvalues is a numerical artifact which vanishes in the continuum limit. This feature is a common occurrence in the study of spectral stability, and arises from the numerical model being ill-conditioned; that the soliton, which is strictly speaking a solution in free space, is discretised and truncated in the numerical picture. 

\begin{figure}[t]
\includegraphics[width=8.6cm]{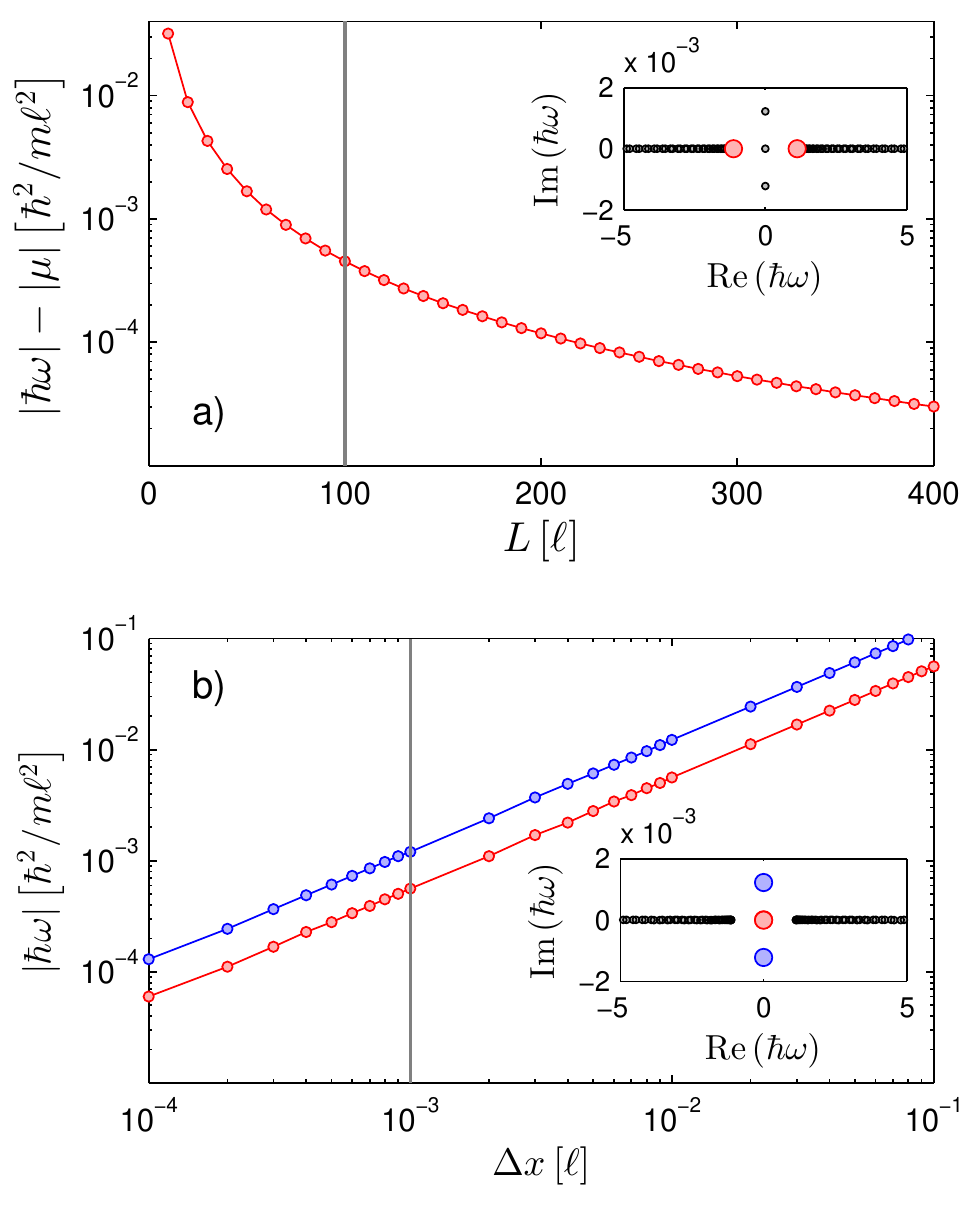}
\caption{(colour online). Numerical convergence of the Bogoliuvbov de-Gennes eigenvalues in the continuum limit. (a) The continuous state band-edge $(q=0)$ eigenvalue compared to the chemical potential for increasing domain length, and (b) discrete eigenvalues for decreasing domain spacing. The soliton parameters are fixed at $g_{\mathrm{1D}}m\ell /\hbar ^{2}=-1$, $vm\ell /\hbar =1$, and $a_{1}/\hbar =1$, with the grey line of each plot corresponding to the spectrum shown in the respective inset with the eigenvalues colour-coded.}
\label{image_eigs_test}
\end{figure}

To resolve this discrepancy, we define the numerical domain of the soliton as $\left[-L/2,L/2\right]$, with length $L$ and spacing $\Delta x$, provided $L>b$. Then, in the continuum limit, where $L\rightarrow\infty$ and $\Delta x\rightarrow 0$, it is expected that the numerical eigenvalue problem will become well-conditioned and match the analytical results. To demonstrate this, we numerically solve Eq.~\eqref{bogoeqn_1} in each limit independently for fixed $\tilde{g}_\mathrm{1D}$ and plot the behaviour of the eigenvalues in Fig.~\ref{image_eigs_test}(a) and Fig.~\ref{image_eigs_test}(b) respectively. 

In both cases, we find that the eigenspectrum converges to the exact values, with the $q=0$ continuous state approaching $\vert\hbar\omega\vert=\vert\mu\vert$ (as per Eq.~\eqref{eigs_cont}) when the domain length is increased, and both pairs of discrete eigenvalues converging linearly to $\hbar\omega=0$ when the domain spacing decreases. Note, that the discrete eigenvalues do not need to be considered in the former case as they are localised to  the width of the soliton and will therefore be invariant to variations in $L$, provided $L>b$ by an order of magnitude. For these reasons, we may conclude that for the fixed chemical potential in this example, a reasonably accurate solution for the eigenspectrum can be obtained with a modest domain length of $L=100$ and spacing $\Delta x=10^{-3}$, corresponding each to a error of $\approx 10^{-3}$.

\section{Analytical methods}\label{appen_hyper}
In this appendix, we show how to construct the zero-eigenvalue solutions of the Bogoliubov-de Gennes equations. As this method uses standard techniques which are well documented in the literature \cite{book_flugge,book_LL,book_DJ}, we present the following calculations solely for the sake of completeness. 

We return to the zero-eigenvalue problem described by Eqs.~\eqref{f_eqn} and \eqref{g_eqn}, which contain a homogeneous and an inhomogeneous eigenvalue problem for the eigenvector components $u(x)$ and $v(x)$ respectively. The general solution 
\begin{equation}
 \zeta= 
 \begin{pmatrix}
u_c \\
 v_c
 \end{pmatrix}+ \begin{pmatrix}
0 \\
 v_p
 \end{pmatrix}=\zeta_c+\zeta_p,
\end{equation}
will therefore be composed of a complementary solution $\zeta_c$ for the reduced homogeneous system and a particular solution $\zeta_p$ to be solved for successively.
\subsection{Complementary solution}
By introducing the soliton width as a scaling parameter, we write the homogeneous system as
\begin{equation}\label{pt_1}
\left[-\frac{d^2}{d\chi^2}-\ell\left(\ell+1\right)\sech^2\chi\right]\zeta_c=E \zeta_c,
\end{equation}
with $\chi=x/b$, $\ell\in\mathbb{Z}^+_0$, and dimensionless eigenvalue $E$.  The potential function appearing in Eq.~\eqref{pt_1} is commonly referred to as a modified P\"{o}schl-Teller potential, and has been studied in the context of reflectionless scattering \cite{lekner_2007,var_1} and supersymmetry \cite{book_super,diaz_1999}. As this potential is attractive, and converges to zero when $x\rightarrow\pm\infty$, the corresponding eigenspectrum will consist of two sets: a bound-state spectrum for $E<0$, and scattering states with $E>0$. For our purposes, we will only be concerned with the bound-state spectrum.

To proceed we follow the method outlined by Fl\"{u}gge \cite{book_flugge}, in which we seek to transform Eq.~\eqref{pt_1} into hypergeometric form by introducing the change of variables $y=\cosh^2\chi$. The resulting transformed differential equation takes the form
\begin{equation}
y\left(1-y\right)\frac{d^2\zeta_c}{dy^2}+\left[\frac{1}{2}-y\right]\frac{d\zeta_c}{dy}-\left[\frac{\ell\left(\ell+1\right)}{4y}+\frac{E}{4}\right]\zeta_c=0.
\end{equation}
Then, by further setting
\begin{equation}
\zeta_c=wy^{\left(\ell+1\right)/2},
\end{equation}
we arrive at the hypergeometric differential equation
\begin{equation}\label{hyper_1}
y\left(1-y\right)\frac{d^2w}{dy^2}+\left[\gamma-\left(\alpha+\beta+1\right)y\right]\frac{dw}{dy}-\alpha\beta w=0,
\end{equation}
with the abbreviations $\alpha=(\ell+1-i\sqrt{E}\;)/2$, $\beta=(\ell+1+i\sqrt{E}\;)/2$, and $\gamma=\ell+3/2$. For the domain $0\leq\vert x\vert\leq\infty\rightarrow 1\leq y\leq\infty$, the general solution around the singular point $y=1$ is given by \cite{book_math}
\begin{equation}\label{hyper_sol}
\begin{split}
\zeta_c/&y^{\left(\ell+1\right)/2}=A_{\ 2}F_1\left(\alpha,\beta;\gamma^\prime;1-y\right)\\&+\left(1-y\right)^{-\gamma^\prime+1}B_{\ 2}F_1\left(\gamma-\alpha,\gamma-\beta;2-\gamma^\prime;1-y\right)
\end{split},
\end{equation}
where $A$ and $B$ are arbitrary constants with $\gamma^\prime=\alpha+\beta-\gamma+1$. 

As we require the solutions described by Eq.~\eqref{hyper_sol} to be square-integrable for $E<0$, we can derive an expression for the corresponding eigenvalues by studying the asymptotic behaviour of the solutions. Introducing the asymptotic expression $y=\cosh^2\chi\sim e^{2\vert \chi\vert}/4$ and using Kummer's solutions \cite{book_math}, one can write the independent solutions as
\begin{equation}
\begin{split}
\lim_{\vert x\vert\to\infty} A_{\ 2}F_1\sim y^{-\alpha}&\frac{\Gamma\left(\gamma^\prime\right)\Gamma\left(\beta-\alpha\right)}{\Gamma\left(\beta\right)\Gamma\left(\beta-\gamma+1\right)}\\
&+y^{-\beta}\frac{\Gamma\left(\gamma^\prime\right)\Gamma\left(\alpha-\beta\right)}{\Gamma\left(\alpha\right)\Gamma\left(\alpha-\gamma+1\right)}
\end{split},
\end{equation}
and
\begin{equation}
\begin{split}
\lim_{\vert x\vert\to\infty} B_{\ 2}F_1\sim &\left(1-y\right)^{-2}y^{\beta-\gamma}\frac{\Gamma\left(2-\gamma^\prime\right)\Gamma\left(\beta-\alpha\right)}{\Gamma\left(1-\alpha\right)\Gamma\left(\gamma-\alpha\right)}\\
&+\left(1-y\right)^{-2}y^{\alpha-\gamma}\frac{\Gamma\left(2-\gamma^\prime\right)\Gamma\left(\alpha-\beta\right)}{\Gamma\left(1-\beta\right)\Gamma\left(\gamma-\beta\right)}
\end{split}.
\end{equation}
The pair of asymptotic forms described above converge provided the ratio of $\Gamma$-functions vanishes. Therefore we require $\beta-\gamma+1=-n$ and $1-\alpha=-n$, for $n\in\mathbb{Z}^+_0$ provided $E\leq 0$. The resulting expression for the eigenvalues can then be determined iteratively, and takes the form
\begin{equation}
E=-\left(\ell-n\right)^2,
\end{equation}
with the constraint $0\leq n\leq \ell-1$.

To continue, it is instructive to consider specific values of $\ell$ and $E$, to obtain the solution as required. Solving for $f_c$ in Eq.~\eqref{pt_1} with $\ell=1$ and $E=-1$, we find the complementary solution
\begin{equation}
u_c=A\sech\chi+i\frac{B}{2}\left(\chi\sech\chi+\sinh\chi\right),
\end{equation}
using the hypergeometric identities
\begin{equation}
_{\ 2}F_1\left(1/2,3/2;1/2;-\sinh^2\chi\right)=\sech^3\chi,
\end{equation}
and
\begin{equation}
\begin{split}
_{\ 2}F_1\left(2,1;3/2;-\sinh^2\chi\right)&=\\
\frac{1}{2}\sech^2\chi &\left(\chi\csch\chi\sech\chi+1\right).
\end{split}
\end{equation}
Then in the same manner for $g_c$ in Eq.~\eqref{pt_1} with $\ell=2$ and $E=-1$, we find the complementary solution
\begin{equation}
\begin{split}
v_c&=\frac{A}{2}\sech\chi\left(3-3\chi\tanh\chi-\cosh^2\chi\right)\\&
+iB\tanh\chi\sech\chi,
\end{split}
\end{equation}
with
\begin{equation}
\begin{split}
_{\ 2}F_1&\left(1,2;1/2;-\sinh^2\chi\right)=\\
&\frac{3}{2}\sech^4\chi\left(1-\chi\tanh\chi-\left(1/3\right)\cosh^2\chi\right),
\end{split}
\end{equation}
and
\begin{equation}
_{\ 2}F_1\left(5/2,3/2;3/2;-\sinh^2\chi\right)=\sech^5\chi.
\end{equation}
This completes the solution of the homogeneous problem.
\subsection{Particular solution}
With the complementary solution derived, we may now proceed in solving for the particular solution of Eq.~\eqref{g_eqn} using the method of variation of parameters. Labelling the pair of fundamental solutions of the homogeneous problem as
\begin{equation}
v_1=\sech\chi\left(1-\chi\tanh\chi-\cosh^2\chi/3\right),
\end{equation}
and
\begin{equation}
v_2=\tanh\chi\sech\chi,
\end{equation}
we write the Wronskian relation
\begin{equation}
W\left(v_1,v_2\right)=v_1\frac{dv_2}{dx}-v_2\frac{dv_1}{dx}=2/3b.
\end{equation}
The particular solution can then be obtained by direct integration
\begin{equation}\label{var_of_par}
v_p=-v_1\int dx\;\frac{v_2\;h\left(x\right)}{W\left(v_1,v_2\right)}+v_2\int dx\;\frac{v_1\;h\left(x\right)}{W\left(v_1,v_2\right)},
\end{equation}
with $h(x)=-2\mathcal{J}_0u_c$ corresponding to the inhomogeneous part of Eq.~\eqref{g_eqn}. As the expressions derived from Eq.~\eqref{var_of_par} are often quite cumbersome, we refer the reader to the solutions presented in Sec.~\ref{section_ana}.

\providecommand{\noopsort}[1]{}\providecommand{\singleletter}[1]{#1}%

\end{document}